\documentclass[12 pt]{article}%

\usepackage{setspace}
\usepackage{amsfonts}
\usepackage{color}
\usepackage[usenames,dvipsnames]{xcolor}
\usepackage{fullpage}
\usepackage{subfigure}
\usepackage{amsmath}
\usepackage{amsthm}
\usepackage{comment}
\usepackage{graphicx}
\usepackage{amssymb,epsfig}
\usepackage[authoryear]{natbib}
\usepackage{amssymb}
\usepackage{authblk}%
\usepackage[font=footnotesize,labelfont=bf]{caption}

\newtheorem{alg}{Algorithm}

\newtheorem{ex}{Example}

\begin{document}

\title{Parameter Estimation in Hidden Markov Models with Intractable Likelihoods Using Sequential Monte Carlo}
\author{Sinan Y{\i}ld{\i}r{\i}m\thanks{School of Mathematics, University of Bristol, UK, \texttt{s.yildirim@bristol.ac.uk}}, Sumeetpal S. Singh\thanks{Department of Engineering, University of Cambridge, UK, \texttt{sss40@cam.ac.uk}}, Thomas Dean\thanks{\texttt{thomas.dean@cantab.net}}, Ajay Jasra\thanks{Department of Statistics and Applied Probability, National University of Singapore, \texttt{staja@nus.edu.sg}}}

\date{\today}
\maketitle

\begin{abstract}
We propose sequential Monte Carlo based algorithms for maximum likelihood estimation of the static parameters in hidden Markov models with an intractable likelihood using ideas from approximate Bayesian computation. The static parameter estimation algorithms are gradient based and cover both offline and online estimation. We demonstrate their performance by estimating the parameters of three intractable models, namely the $\alpha$-stable distribution, \textit{g-and-k} distribution, and the stochastic volatility model with $\alpha$-stable returns, using both real and synthetic data.

\textbf{Key words:} hidden Markov models, maximum likelihood estimation, approximate Bayesian computation, intractable likelihood, sequential Monte Carlo.
\end{abstract}

\section{Introduction} \label{sec: Introduction}

The hidden Markov model (HMM) is an important statistical model used in many fields including bioinformatics (e.g.\ \citet{Durbin_et_al_1998}), econometrics (e.g.\ \citet{Kim_et_al_1998}) and population genetics (e.g.\ \citet{Felsenstein_and_Churchill_1996}); see \citet{Cappe_et_al_2005} for a recent overview. A HMM is comprised of a latent  process $\{ X_{t} \}_{t \geq 1}$ and an observed process $\{ Y_{t} \}_{t \geq 1}$. The latent process is a Markov chain with an initial density $\eta_{\theta}$ and the transition density $f_{\theta}$, i.e.\
\begin{align} \label{eq: law of hidden state}
X_{t}  \in \mathcal{X} \subseteq \mathbb{R}^{d_{x}}, \quad X_{1} \sim \eta_{\theta}(\cdot), \quad X_{t} | (X_{1:t-1} = x_{1:t-1}) \sim f_{\theta}(\cdot | x_{t-1}), \quad t \geq 2.
\end{align}
It is assumed that $\eta_{\theta}(x)$ and $f_{\theta}(x | x^{\prime})$ are densities on $\mathcal{X}$ with respect to a dominating measure denoted generically as $dx$. The observation at time $t$ is conditionally independent of all other random variables given $X_{t} = x_{t}$ and its conditional observation density is $g_{\theta}(\cdot | x_{t})$ on $\mathcal{Y}$ with respect to the dominating measure $dy$, i.e.\
\begin{align} \label{eq: law of observation}
Y_{t}  \in \mathcal{Y} \subseteq \mathbb{R}^{d_{y}}, \quad Y_{t} |  \{ x_{i} \}_{i \geq 1}, \{ y_{i} \}_{i \geq 1, i \neq t}  \sim g_{\theta}( \cdot | x_{t}), \quad t \geq 1.
\end{align}
The law of the HMM is parametrised by a vector $\theta$ taking values in some compact subset $\Theta$ of the Euclidean space $\mathbb{R}^{d_{\theta}}$.

In this paper we focus on HMMs where the probability density $g_{\theta}(y | x)$ of the observations is \textit{intractable}. By intractable we mean that $g_{\theta}(y | x)$ cannot be evaluated (or it is computationally prohibitive to calculate). However, we are able to generate samples from $g_{\theta}(\cdot | x)$ despite its intractability. 

We will denote the actual observed random variables of the HMM as $\hat{y}_{1}, \hat{y}_{2}, \ldots$ and assume that they are generated by some unknown $\theta^{\ast} \in \Theta$ which is to be estimated. The maximum likelihood estimate of $\theta^{\ast}$ given $\hat{y}_{1:n}$ is
\begin{equation}
\theta_{\text{ML}} = \arg \max_{\theta \in \Theta} p_{\theta}(\hat{y}_{1:n}), \nonumber
\end{equation}
where $p_{\theta}(\hat{y}_{1:n})$ is the probability density, or the \textit{likelihood}, of the observations $\hat{y}_{1:n}$, and from \eqref{eq: law of hidden state}-\eqref{eq: law of observation}, is given by
\begin{equation} \label{eq: true likelihood}
p_{\theta}(\hat{y}_{1:n}) = \int_{\mathcal{X}^{n}} \eta_{\theta}(x_{1}) g_{\theta}(\hat{y}_{1} | x_{1}) \left[ \prod_{t = 2}^{n} f_{\theta}(x_{t} | x_{t-1}) g_{\theta}(\hat{y}_{t} | x_{t} ) \right] dx_{1:n}. 
\end{equation}
Even when $\mathcal{X}$ is a finite set, $p_{\theta}(\hat{y}_{1:n})$ cannot be evaluated because $g_{\theta}(y | x)$ is intractable. There is a sizeable literature on the use of sequential Monte Carlo (SMC) methods, also known as particle filters, to evaluate the gradient of $p_{\theta}(\hat{y}_{1:n})$ with respect to $\theta$, which is subsequently used to compute its maximiser, see e.g.\ the review in \citet{Kantas_et_al_2009}. However, these methods require a \emph{tractable} $g_{\theta}(y | x)$ and they are not directly applicable when this density is intractable. We thus propose new SMC based maximum likelihood estimation (MLE) algorithms to fill this void. We handle the intractable $g_{\theta}(y | x)$ by drawing on ideas from approximate Bayesian computation (ABC), an inference technique initially developed for Bayesian models with an intractable likelihood, see \citet{Marin_et_al_2012} for a recent review. Our static parameter estimation algorithms are gradient based and cover both offline (or batch) and online estimation.  

Recently \citet{Ehrlich_et_al_2013} have proposed a gradient based MLE algorithm for HMMs with an intractable observation density $g_{\theta}(y | x)$. The authors estimate the gradient of the likelihood in \eqref{eq: true likelihood} using a finite difference approximation which ultimately relies on estimates of $p_{\theta}(\hat{y}_{1:n})$ only, which itself is calculated using SMC. The major advantage of our method over that of \citet{Ehrlich_et_al_2013} is that we characterise the gradient of the log likelihood directly, by using available information on how the intractable $g_{\theta}(y | x)$ is simulated from, and subsequently approximate it using SMC, thus avoiding the added error of a finite difference approximation. Our online MLE algorithm is asymptotically unbiased (as our numerical results indicate) as the number of particles increases whereas the same cannot be said for \citet{Ehrlich_et_al_2013} due to the finite difference approximation; their numerical results indicate a bias that does not diminish with increasing data, even when $p_{\theta}(\hat{y}_{1:n})$ can be calculated exactly as they illustrate for a linear Gaussian state-space model (see \citet[Figure 2]{Ehrlich_et_al_2013}). Also, as observed from the results in \citet{Ehrlich_et_al_2013}, the variance of the parameter estimates of their recursive MLE algorithm does not diminish with more data while ours does (see the discussion in Section \ref{sec: Gradient ascent}).

The remainder of this paper is organised as follows. The theory that underpins our MLE methodology is detailed in Section \ref{sec: The ABC MLE approach for parameter estimation} and in Section \ref{sec: Implementing ABC MLE with SMC} we describe its SMC implementation. Numerical examples using both simulated and real data sets are given in Section \ref{sec: Numerical examples}. The numerical work covers three intractable models, namely the $\alpha$-stable distribution, \textit{g-and-k} distribution, and the stochastic volatility model with $\alpha$-stable returns. Finally, Section \ref{sec: Discussion} provides a discussion of other possible methods for parameter estimation in HMMs when both state and observation densities intractable. 

\section{The ABC MLE approach for parameter estimation} \label{sec: The ABC MLE approach for parameter estimation}

The particle filter sequentially approximates the sequence of posterior densities $\{ p_{\theta}(x_{1:t} | Y_{1:t} = \hat{y}_{1:t}) \}_{t \geq 1}$ of the HMM $\{ X_{t}, Y_{t} \}_{t \geq 1}$ using a weighted discrete distribution with $N$ support points for $X_{1:t}$ which are called particles. At each time $t$, the particles are resampled according to their current weights, and then the resampled particles are propagated independently of each other using a proposal transition density $r_{\theta}(x_{t+1} | x_{t})$. The particles are then reweighed to correct for the discrepancy between $p_{\theta}(x_{1:t+1} | Y_{1:t+1} = \hat{y}_{1:t+1})$ and the law of the proposed particles which is $p_{\theta}(x_{1:t} | Y_{1:t} = \hat{y}_{1:t}) r_{\theta}(x_{t+1} | x_{t})$. This is standard importance sampling and the assumption in the weight correction step is that the law of each resampled particle at time $t$ is $p_{\theta}(x_{1:t} | Y_{1:t} = \hat{y}_{1:t})$, which is an erroneous but progressively correct as $N$ is increased  \citep{Del_Moral_2004,  Crisan_and_Doucet_2002, Chopin_2002}. In the implementation of the particle filter the normalising constants of the sequence of target posteriors are not needed but calculating the new weights requires $g_{\theta}(\hat{y} | x)$ to be tractable. It was shown by \citet{Del_Moral_2004} that the weights of the particle approximation of $\{ p_{\theta}(x_{1:t} | Y_{1:t} = \hat{y}_{1:t}) \}_{t \geq 1}$ can be used to obtain an unbiased estimate of the likelihoods $\{ p (Y_{1:t} = \hat{y}_{1:t}) \}_{t \geq 1}$. See the Appendix for an example code for a particle filter.

\citet{Jasra_et_al_2012} consider the problem of constructing an SMC approximation of the \emph{filter} $p_{\theta}(x_{t} | Y_{1:t} = \hat{y}_{1:t})$, which is the marginal of the particle approximation for $p_{\theta}(x_{1:t} | Y_{1:t} = \hat{y}_{1:t})$, for a HMM with an intractable observation density $g_{\theta}(y | x)$. Since it is not possible to calculate the weights of the particle filter for such an HMM where $g_{\theta}(y | x)$ is intractable, they propose a particle filter approximation for the extended HMM $\{ (X_{t}, Y_{t}), Y^{\epsilon}_{t} \}_{t \geq 1}$ where the joint process $\{ X_{t}, Y_{t} \}_{t \geq 1}$, which is now the latent process of the extended HMM, is defined by \eqref{eq: law of hidden state} and \eqref{eq: law of observation} and the new sequence $\{ Y_{t}^{\epsilon} \}_{t \geq 1}$ is
\begin{equation}
Y^{\epsilon}_{t} = Y_{t} + \epsilon V_{t}, \quad V_{t} \sim^{\text{i.i.d.}} \text{Unif}(B_{0}^{1}), \quad t \geq 1,
\label{eq:noisyObs}
\end{equation}
where $B_{y}^{r}$ denotes the ball of radius $r > 0$ centred at $y \in \mathbb{R}^{d_{y}}$ and $\text{Unif}(B)$ is the uniform distribution over the set $B$. Then, the density 
\[
p_{\theta^{\ast}}(x_{t} | Y^{\epsilon}_{1:t} = \hat{y}_{1:t})
\]
of the extended HMM is regarded as an approximation for $p_{\theta^{\ast}}(x_{t} | Y_{1:t} = \hat{y}_{1:t})$ where $\epsilon > 0$ reflects the error of the approximation and this error diminishes as $\epsilon \rightarrow 0$; see also \citet{Calvet_and_Czellar_2012} and \citet{Martin_et_al_2012} for theoretical results on this approximation. Note that $p_{\theta^{\ast}}(x_{t} | Y^{\epsilon}_{1:t} = \hat{y}_{1:t})$ does not coincide with $p_{\theta^{\ast}}(x_{t} | Y_{1:t} = \hat{y}_{1:t})$ because $\hat{y}_{1:t}$ obeys the law \eqref{eq: law of hidden state}-\eqref{eq: law of observation} and not \eqref{eq:noisyObs}. \citet{Jasra_et_al_2012} remark that $p_{\theta^{\ast}}(x_{t} | Y^{\epsilon}_{1:t} = \hat{y}_{1:t})$ is the ABC approximation for the filter of a HMM. Furthermore, they show it is straightforward to approximate $p_{\theta^{\ast}}(x_{t} | Y^{\epsilon}_{1:t} = \hat{y}_{1:t})$ with a bootstrap particle filter.

Consider now the extended HMM $\{ (X_{t}, Y_{t}), Y^{\epsilon}_{t} \}_{t \geq 1}$  specified by \eqref{eq: law of hidden state}, \eqref{eq: law of observation} and \eqref{eq:noisyObs} and let  $p_{\theta}(Y^{\epsilon}_{1:n} = y_{1:n})$ denote the probability density (or likelihood function) of the process $\{ Y^{\epsilon}_{t} \}_{t \geq 1}$  evaluated at some $y_{1:n} \in (\mathbb{R}^{d_{y}})^n$. (See \eqref{eq: likelihood of the expanded HMM} for the precise expression of this density.) \citet{Dean_et_al_2011} study the theoretical properties of the following maximum likelihood estimate of $\theta^{\ast}$:
\begin{equation}
\theta_{n}^{\epsilon} = \arg \max_{\theta \in \Theta} p_{\theta}(Y_{1:n}^{\epsilon}=\hat{y}_{1:n}). \label{eq: ABC MLE solution}
\end{equation}
They call this procedure ABC MLE. (Note that despite the word `Bayesian' in ABC, the procedure is not Bayesian.) The bootstrap particle filter of \citet{Jasra_et_al_2012} provides an unbiased SMC approximation of the likelihood $p_{\theta}(Y_{1:n}^{\epsilon} = \hat{y}_{1:n})$ and this likelihood may be maximised by evaluating the approximation over a grid of values for $\theta$. This, however, is clearly not practical as the dimension of $\theta$ increases, has no straightforward extension for recursive estimation and is not an accurate convergent method. 

It was shown in \citet{Dean_et_al_2011} that the ABC MLE \eqref{eq: ABC MLE solution} leads to a biased estimate of the parameter vector $\theta^{\ast}$ in the sense that as $n \to \infty$, $\theta_{n}^{\epsilon}$ will converge to some point $\theta^{\ast, \epsilon} \neq \theta^{\ast} \in \Theta$, and that this bias can be made arbitrarily small by choosing a sufficiently small value of $\epsilon$, i.e.\ $\theta^{\ast, \epsilon} \rightarrow \theta^{\ast}$ as $\epsilon \rightarrow 0$. The bias of ABC MLE is due to the fact the observed sequence $\hat{y}_1,\hat{y}_2,\ldots$ is the outcome of the law \eqref{eq: law of observation} for $\theta = \theta^{\ast}$ and not \eqref{eq:noisyObs}. \citet{Dean_et_al_2011} suggest removing the bias of $\theta_{n}^{\epsilon}$ in \eqref{eq: ABC MLE solution} by adding noise to the real data and then computing the maximum likelihood estimate, i.e. let $v_{1}, \ldots, v_{n}$ be a realisation of i.i.d.\ samples from $\text{Unif}(B_{0}^{1})$ and let  
\begin{equation} \label{eq: noisy ABC data}
y^{\epsilon}_{t} = \hat{y}_{t} + \epsilon v_{t}, \quad 1 \leq t \leq n.
\end{equation}
Note that the noisy data $y^{\epsilon}_{1:n}$ now obeys the law of $\{ Y^{\epsilon}_{t} \}_{t \geq 1}$ when $\theta = \theta^{\ast}$. Therefore, the procedure
\begin{equation}
\theta_{n}^{\epsilon} = \arg \max_{\theta \in \Theta} p_{\theta}(Y_{1:n}^{\epsilon} = y^{\epsilon}_{1:n}) \label{eq: noisy ABC MLE solution}
\end{equation}
can now produce a consistent estimator of the parameter vector $\theta^{\ast}$ as $n \rightarrow \infty$. This result proved by \citet{Dean_et_al_2011} can be interpreted as the frequentist equivalence of Wilkinson's observation that the ABC posterior distribution is exact under the assumption of model error \citep{Wilkinson_2013}.

Finally, \citet{Dean_et_al_2011} also remark that the use of other types of noise in \eqref{eq:noisyObs} is possible without compromising the asymptotics of noisy ABC MLE, i.e. 
\begin{equation}
Y^{\epsilon}_{t} = Y_{t} + \epsilon V_{t}, \quad V_{t} \sim^{\text{i.i.d.}} \kappa, \quad t \geq 1,
\label{eq:noisyObs2}
\end{equation}
where $\kappa$ is a smooth centred density.  (Accordingly, noisy ABC MLE is performed with the noise corrupted observations \eqref{eq: noisy ABC data} where now $v_{i}$ are realisations of i.i.d.\ samples from $\kappa.$) As we show, a continuously differentiable $\kappa$ is important for the development of practical gradient based MLE techniques. In this work we choose $\kappa$ to be the probability density of zero-mean unit-variance Gaussian random variable. Other choices are possible (but not investigated) and our framework would still be applicable.
 
We remark that although the theoretical basis for ABC MLE was established in \citet{Dean_et_al_2011},  the authors do not propose a practical methodology for implementing ABC MLE in their work; this is indeed an important void to be filled. In this paper we demonstrate how, by using ideas from  \citet{Poyiadjis_et_al_2011}, both batch and online versions of noisy ABC MLE can be implemented with SMC.

\section{Implementing ABC MLE with SMC} \label{sec: Implementing ABC MLE with SMC}

We assume that for all $(x, \theta) \in \mathcal{X} \times \Theta$ there exist a distribution on some auxiliary space $\mathcal{U}$ with a tractable density $\nu_{\theta}( \cdot | x) $ with respect to $du$ and a function $\tau_{\theta}: \mathcal{U} \times \mathcal{X} \rightarrow \mathcal{Y}$  such that one can sample from $g_{\theta}(\cdot | x)$ by first sampling $ U \in \mathcal{U}$ from $\nu_{\theta}( \cdot | x) $ and then applying the transformation $U \rightarrow \tau_{\theta}(U, x)$; i.e.\ the law of $\tau_{\theta}(U, x)$ is $g_{\theta}(\cdot | x)$. From this it follows that the process $\{Y^{\epsilon}_{t}\}_{t \geq 1}$ in \eqref{eq:noisyObs2} can be equivalently generated as 
\begin{equation}
Y^{\epsilon}_{t} = \tau_{\theta} ( U_{t}, X_{t} ) + \epsilon V_{t}, \quad V_{t} \sim^{\text{i.i.d.}} \kappa, \quad  t \geq 1. \label{eq:noisyObs3}
\end{equation}
where $\left\{ X_{t} \right\}_{t \geq 1}$ is the hidden state of the original HMM given by \eqref{eq: law of hidden state} and $U_{t} \sim \nu_{\theta}(\cdot | X_{t})$ for all $t$. We will implement SMC based MLE for the following HMM: Let  $\{ Z_{t} := \left( X_{t}, U_{t}\right) \}_{t \geq 1}$ be the latent process and $\left\{ Y^{\epsilon}_{t} \right\}_{t \geq 1}$ in \eqref{eq:noisyObs3} be the observation process. The initial and transition densities for $\{ Z_{t} \}_{t \geq 1}$ (with respect to the dominating measure $dz = dx du$) and the observation density of $\{ Y_{t}^{\epsilon} \}_{t \geq 1}$ (with respect to the Lebesgue measure on $\mathbb{R}^{d_{y}}$) are
\begin{align}
\pi_{\theta}(z) = \eta_{\theta}(x) \nu_{\theta}(u | x), \quad q_{\theta}(z' | z) = f_{\theta}(x' | x) \nu_{\theta}(u' | x'), \quad h_{\theta}^{\epsilon}(y | z) = \frac{1}{\epsilon} \kappa \left( \frac{y - \tau_{\theta}(z)}{\epsilon} \right) \label{eq: law of expanded HMM}
\end{align}
where $z = (x, u)$ and $z' = (x', u')$. The density of the observed process $Y^{\epsilon}_{1:n}$ of this HMM evaluated at some $y_{1:n}$ is
\begin{align} \label{eq: likelihood of the expanded HMM}
p_{\theta}(y_{1:n}) := \int_{\mathcal{Z}^{n}} \pi_{\theta}(z_{1}) h_{\theta}^{\epsilon}(y_{1} | z_{1})  \left[ \prod_{t = 2}^{n} q_{\theta}(z_{t} | z_{t-1}) h_{\theta}^{\epsilon}(y_{t} | z_{t}) \right] d z_{1:n}.
\end{align}
where $\mathcal{Z} = \mathcal{X} \times \mathcal{U}$. Note that $p_{\theta}(\cdot)$ in \eqref{eq: likelihood of the expanded HMM} is indeed the likelihood function $p_\theta(Y_{1:n}^\epsilon = \cdot)$ to be maximised with respect to $\theta$ in ABC MLE in Section \ref{sec: The ABC MLE approach for parameter estimation}; see \eqref{eq: ABC MLE solution} and \eqref{eq: noisy ABC MLE solution}. Moreover all the densities declared in \eqref{eq: law of expanded HMM} are tractable and differentiable functions of $\theta$ (provided that $f_{\theta}$, $\nu_{\theta}$, and $\tau_{\theta}$ are differentiable with respect to $\theta$).

Henceforth, we will work exclusively with the HMM $\{ Z_{t},Y_t^\epsilon\}_{t \geq 1}$ defined in \eqref{eq: law of expanded HMM}. As discussed before, we corrupt the real measurements $\hat{y}_{1},\hat{y}_{2}, \ldots$ with a single realisation of independent samples $v_{1}, v_{2} \ldots$ from a $\theta$-independent probability density $\kappa$, i.e.\
\[ 
y_{i}^{\epsilon} = \hat{y}_{i} + \epsilon v_{i},
\] 
to obtain a realisation of the observed process of  the HMM $\{ Z_{t}, Y_{t}^{\epsilon} \}_{t \geq 1}$.

\subsection{Gradient ascent} \label{sec: Gradient ascent}
One well known MLE algorithm is the following iterative gradient ascent  method which updates the parameter estimate $\theta_{j}$ using the rule
\begin{equation}
\theta_{j} = \theta_{j-1} + \gamma_{j} \nabla \log p_{\theta_{j-1}}(y^{\epsilon}_{1:n}) \label{eq: gradient MLE update}
\end{equation}
where $\theta_{0} \in \Theta$ is an arbitrary initial estimate. Here $\{ \gamma_{j} \}_{j \geq 1}$ is a sequence of step-sizes satisfying the constraints $\sum_{j \geq 1} \gamma_{j} = \infty$ and $\sum_{j \geq 1} \gamma_{j}^{2} \leq \infty$ so as to ensure that the algorithm converges to a local maximum of $\log p_{\theta}(y^{\epsilon}_{1:n})$. The term $\nabla \log p_{\theta}(y^{\epsilon}_{1:n})$ is shorthand for the $\mathbb{R}^{d_{\theta}}$-valued vector
\[
\nabla \log p_{\theta}(y^{\epsilon}_{1:n}):= \frac{\partial \log p_{\theta}(y^{\epsilon}_{1:n})} {\partial \theta},
\]
which is also called the score vector, and is given by Fisher's identity (e.g.\ see \citet{Cappe_et_al_2005})
\begin{align}
\nabla \log p_{\theta}(y^{\epsilon}_{1:n}) = \int_{\mathcal{Z}^{n}} \left[ \sum_{t = 1}^{n} \nabla \log q_{\theta}(z_{t} | z_{t-1}) + \nabla \log h^{\epsilon}_{\theta}(y^{\epsilon}_{t} | z_{t}) \right]  p_{\theta}(z_{1:n} | y^{\epsilon}_{1:n}) d z_{1:n}
\label{eq:score}
\end{align}
with the convention that $q_{\theta}(z_{1} | z_{0}) = \pi_{\theta}(z_{1}) = \eta_{\theta}(x_{1}) \nu_{\theta}(u_{1} | x_{1})$. Note that the method in \eqref{eq: gradient MLE update} uses the whole data set $y_{1:n}^{\epsilon}$ at every parameter update step, which makes it a batch method. An alternative to it is the following online gradient ascent method which updates the parameter estimate every time a new data point is received
\begin{align}
\theta_{n} &= \theta_{n-1} + \gamma_{n} \nabla \log p_{\theta_{n-1}}(y^{\epsilon}_{n} | y^{\epsilon}_{1:n-1}),
\label{eq: recursive gradient MLE update}
\end{align}
where 
\begin{align} 
\nabla \log p_{\theta_{n-1}}(y^{\epsilon}_{n} | y^{\epsilon}_{1:n-1}) = \nabla \log p_{\theta_{n-1}}(y^{\epsilon}_{1:n}) - \nabla \log p_{\theta_{n-1}}(y^{\epsilon}_{1:n-1}). \label{eq: gradient of the incremental log-likelihood}
\end{align}
While the subscript $\theta_{n-1}$ indicates that $\nabla \log p_{\theta}(y^{\epsilon}_{n} | y^{\epsilon}_{1:n-1})$ is evaluated at $\theta = \theta_{n-1}$, 
a necessary requirement for a truly online implementation is that the previous values of $\theta$ estimates (i.e.\ other than $\theta_{n-1}$) are also used in the evaluation of $\nabla \log p_{\theta_{n-1}}(y^{\epsilon}_{n} | y^{\epsilon}_{1:n-1})$ \citep{Le_Gland_and_Mevel_1997}.

It is important to note that, for both batch \eqref{eq: gradient MLE update} and online method \eqref{eq: recursive gradient MLE update}, we require that the transition density of $\{ Z_{t} \}_{t \geq 1}$ be tractable and differentiable with respect to $\theta$, which is precisely why we propose to work with $\{ Z_{t}, Y^{\epsilon}_{t} \}_{t \geq 1}$ rather than $\{ (X_{t}, Y_{t}), Y^{\epsilon}_{t} \}_{t \geq 1}$ whose state transition density contains the intractable $g_{\theta}$. (We discuss suitable alternatives when the state transition density is intractable in Section \ref{sec: Other MLE methods for HMMs with an intractable density}.)

It is apparent from \eqref{eq: gradient MLE update} and \eqref{eq: recursive gradient MLE update} that an SMC implementation of these MLE algorithms hinges on the availability of a particle approximation of the score in \eqref{eq:score}. \citet{Poyiadjis_et_al_2011} discuss two methods to estimate the score using the SMC approximation of the full posterior $p_{\theta} (z_{1:n} | y^{\epsilon}_{1:n})$. One method is nothing more than the substitution of the law $p_{\theta} (z_{1:n} | y^{\epsilon}_{1:n})$ in \eqref{eq:score} with its particle approximation and has a cost, like the particle filter itself, which is $\mathcal{O}(N)$. We will refer to this estimate of the gradient as the $\mathcal{O}(N)$ method \citep[Algorithm 1]{Poyiadjis_et_al_2011}. Due to resampling step of the particle filter there is a lack of unique samples in the particle approximation of $p_{\theta} (z_{1:m} | y^{\epsilon}_{1:n})$ for $m$ much smaller than $n$, which is called particle degeneracy in the literature. \citet{Poyiadjis_et_al_2011} shows that the variance of this $\mathcal{O}(N)$ score estimate, where the variance is computed with respect to the particles being sampled while the observation sequence is held fixed, grows quadratically with time. While this may not be an issue for the batch method in \eqref{eq: gradient MLE update}, it is not suitable for online estimation \eqref{eq: recursive gradient MLE update} since the variance of the resulting estimate of $\nabla \log p_{\theta_{n-1}}(y^{\epsilon}_{n} | y^{\epsilon}_{1:n-1})$ grows linearly with time $n$.

As an alternative to this standard $\mathcal{O}(N)$ score estimate, \citet{Poyiadjis_et_al_2011} propose an $\mathcal{O}(N^2)$ estimate of the score computed using the same particle approximation to $p_{\theta} (z_{1:n} | y^{\epsilon}_{1:n})$ which aims to avoid the particle degeneracy problem mentioned. We will refer to this as the $\mathcal{O}(N^2)$ method \citep[Algorithm 2]{Poyiadjis_et_al_2011}. The authors experimentally show that the variance of the score estimate now grows linearly in time $n$ while the variance of the resulting estimate of $\nabla \log p_{\theta_{n-1}}(y^{\epsilon}_{n} | y^{\epsilon}_{1:n-1})$ is time-uniformly bounded (i.e.\ does not grow); a proof of the latter fact can be found in \citet{Del_Moral_et_al_2011}. Therefore, the SMC implementation of $\nabla \log p_{\theta}(y^{\epsilon}_{n} | y^{\epsilon}_{1:n-1})$ we adopt for online estimation \eqref{eq: recursive gradient MLE update} is the $\mathcal{O}(N^{2})$ method.

Finally, we mention that the score \eqref{eq: gradient MLE update} can also be estimated using a fixed-lag method which would have a computational cost which is $\mathcal{O}(N)$ and a variance which grows linearly in time. However there is the added error introduced by not smoothing beyond a certain lag; see \citet{Kantas_et_al_2009} for a review of static parameter estimation techniques.

\subsection{Controlling the variance of the gradient estimate} \label{sec: Controlling the variance of the gradient estimate}
If the Monte Carlo estimates of the gradient terms have high or infinite variances, we expect failure of the gradient ascent methods. We can stabilise the variance by transforming the observed data, but without compromising the identifiability of the model, and then add noise as discussed in noisy ABC. This approach to stabilising the variance is novel as the issue of infinite variance has not been reported before in the SMC literature.

This issue of the potential for infinite variance (prior to stabilising by adopting a specific transformation) can be perfectly exemplified by the problem of learning the parameters of a distribution from a sequence of i.i.d.\ random variables which we now discuss. Let $\{ Y_{t} \}_{t \geq 1}$ be an i.i.d.\ sequence with an intractable probability density $g_{\theta}(dy)$ on $\mathcal{Y}$. For any $\theta$, assume $Y_{t}$ can be sampled from $g_{\theta}$ by first generating $U_{t} \in \mathcal{U} $ from the density $\nu_{\theta}(du)$ and then followed by the application a certain transformation function $\tau_{\theta}: \mathcal{U} \rightarrow \mathcal{Y}$, i.e.\ the law of $\tau_{\theta}(U_{t})$ is $g_{\theta}$. (The $\alpha$-stable process is generated precisely in this way; see Example \ref{ex: stability alpha stable} below.) We are given a realisation $\hat{y}_{1},\hat{y}_{2}, \dots$ from $\theta^{\ast}$ and the latter is to be estimated. Let $y_{t}^{\epsilon}$ be the noise corrupted observed sequence as in \eqref{eq:noisyObs2}. In the context of the discussion in Section \ref{sec: Implementing ABC MLE with SMC}, the aim is to maximise the likelihood of the noisy observations $y^{\epsilon}_{1:n}$ (generated from the true model $\theta^{\ast}$) using the parametric family of HMMs $\{ U_{t}, Y^{\epsilon}_{t} \}_{t \geq 1}$. Since $\{ U_{t} \}_{t \geq 1}$ are i.i.d.\ the batch \eqref{eq: gradient MLE update} and online \eqref{eq: recursive gradient MLE update} update rules become, respectively,
\begin{align}
\theta_{j} = \theta_{j-1} + \gamma_{j} \sum_{t = 1}^{n} \nabla \log p_{\theta_{j-1}}(y^{\epsilon}_{t}) \quad \text{and}\quad \theta_{n} = \theta_{n-1} + \gamma_{n} \nabla \log p_{\theta_{n-1}}(y^{\epsilon}_{n}). \nonumber
\end{align}
$h_{\theta}^{\epsilon}$ in \eqref{eq: law of expanded HMM} becomes $h_{\theta}^{\epsilon}(y | u) = \frac{1}{\epsilon} \kappa \left( \frac{y - \tau_{\theta}(u)}{\epsilon} \right)$
and 
\begin{align} \label{eq: incremental log-likelihood for i.i.d. case}
\nabla \log p_{\theta}(y^{\epsilon}) &= \int_{\mathcal{Y}} \left[ \nabla \log \nu_{\theta} (u) + \nabla \log h_{\theta}^{\epsilon}(y^{\epsilon} | u) \right] p_{\theta}(u | y^{\epsilon})  du,
\end{align} 
where $p_{\theta}(u | y^{\epsilon}) \propto h_{\theta}^{\epsilon}(y^{\epsilon} | u)  \nu_{\theta} (u).$
Therefore $\nabla \log p_{\theta}(y^{\epsilon}_{n})$ can be estimated using an $N$-sample Monte Carlo approximation to $p_{\theta}(u | y^{\epsilon}_{n})$, e.g.\ with either MCMC or importance sampling. One important point to note about this i.i.d.\ case is that the $\mathcal{O}(N^{2})$ method becomes $\mathcal{O}(N)$.

We now calculate the variance of the Monte Carlo estimate of \eqref{eq: incremental log-likelihood for i.i.d. case} at $\theta = \theta^{\ast}$ given $N$ i.i.d.\ samples from $p_{\theta}(u | y^{\epsilon}_{n}).$ (Note that in the numerical examples we actually use importance sampling to sample from $p_{\theta}(u | y^{\epsilon}_{n})$ but the following calculation is done assuming i.i.d.\ samples are available for illustrative purposes.) Dropping the index $t$, given a noise corrupted measurement $Y^{\epsilon}$ generated from the true model $\theta^{\ast}$, and i.i.d.\ samples $U_{1}, \ldots, U_{N} \sim^{i.i.d.} p_{\theta^\ast}(u | Y^{\epsilon})$, an estimate of $\nabla \log p_{\theta^\ast}(Y^{\epsilon})$ is
\[
\frac{1}{N} \sum_{i = 1}^{N} \frac{1}{\epsilon^{2}} \nabla \tau_{\theta^\ast}(U_{i}) \left[ Y^{\epsilon} - \tau_{\theta^\ast}(U_{i}) \right] + \nabla \log \nu_{\theta^\ast}(U_{i}).
\]
We are interested in the variance of this quantity with respect to the law of $(U_{1:N},Y^{\epsilon})$. We consider the case where $\nabla \log \nu_{\theta^{\ast}}(U)$ has a finite second moment, e.g.\ see the example to follow. Then, the sum above has a finite second moment if and only if $\nabla \tau_{\theta^\ast}(U_{i}) \left[ Y^{\epsilon} - \tau_{\theta^\ast}(U_{i}) \right]$ has a finite second moment with respect to the joint law of $(U_{i}, Y^{\epsilon})$. One can show that 
\begin{align}
\mathbb{E}_{\theta^{\ast}} \left[ \left\{ \nabla \tau_{\theta^{\ast}}(U_{i}) \left[ Y^{\epsilon} - \tau_{\theta^{\ast}}(U_{i}) \right] \right\}^{2} \right] = \epsilon^{2} \mathbb{E}_{\theta^{\ast}} \left[ \left\{ \nabla \tau_{\theta^{\ast}}(U_{i}) \right\}^{2} \right]
\label{eq:infvarexample}
\end{align}
If the second moment of $\nabla \tau_{\theta^{\ast}}$ is infinite (or very high), we may circumvent this instability problem by transforming the \emph{actual} observed process from $\theta^{\ast}$ using a suitable one-to-one function $\psi: \mathcal{Y} \rightarrow \mathcal{Y}_{s}$ prior to adding noise. That is, we replace \eqref{eq:noisyObs2} with the following transformed noise corrupted process 
\begin{equation}  \label{eq: stable noisy observation}
Y^{\epsilon}_{t} = \psi(Y_{t}) + \epsilon V_{t}, \quad V_{t} \sim^{\text{i.i.d.}} \kappa, \quad t \geq 1.
\end{equation}
The conditional density $h_{\theta}^{\epsilon}(y | u)$ becomes
\[
h_{\theta}^{\epsilon}(y | u) = \frac{1}{\epsilon} \kappa \left( \frac{y - \psi [\tau_{\theta}(u)]}{\epsilon} \right)
\]
and the right hand side of \eqref{eq:infvarexample} now is $\epsilon^{2} \mathbb{E}_{\theta^\ast} \left[ \left\{ \nabla \psi(\tau_{\theta^\ast}(U_{i})) \right\}\right]$. In this paper we use $\psi=\tan^{-1}$ throughout, and in the following example we show how \eqref{eq:infvarexample} is infinite but subsequently stabilised with this transformation.

\begin{ex} \label{ex: stability alpha stable} \textbf{(The $\alpha$-stable distribution.)} 
Let $\mathcal{A}(\alpha, \beta, \mu, \sigma)$ denote the $\alpha$-stable distribution. The parameters of the distribution, 
\[
\theta = (\alpha, \beta, \mu, \sigma) \in \Theta = (0, 2] \times [-1, 1] \times \mathbb{R} \times [0, \infty),
\]
represent the shape, skewness, location, and scale respectively. One can generate a random sample from $\mathcal{A}(\alpha, \beta, \mu, \sigma)$ by generating $U = (U_{1}, U_{2})$, where $U_{1} \sim \emph{\text{Unif}}(-\pi/2, \pi/2)$ and $U_{2} \sim \emph{\text{Exp}}(1)$ are independent, and setting 
\[
Y = \tau_{\theta}(U) = \sigma \tau_{\alpha, \beta}(U) + \mu.
\]
The mapping $\tau_{\alpha, \beta}$ is defined as \citep{Chambers_et_al_1976}
\begin{equation}
\tau_{\alpha, \beta}(U) = \begin{cases}
     S_{\alpha, \beta} \frac{\sin \left[ \alpha(U_{1} + B_{\alpha, \beta}) \right]}{\left[ \cos(U_{1}) \right]^{1/\alpha}} \left(\frac{\cos \left[ U_{1} - \alpha (U_{1} + B_{\alpha, \beta}) \right]}{U_{2}} \right)^{(1-\alpha)/\alpha}, & \alpha \neq 1\\
    X = \frac{2}{\pi} \left[ \left( \frac{\pi}{2} + \beta U_{1} \right) \tan U_{1} - \beta \log\left( \frac{U_{2} \cos U_{1}}{\frac{\pi}{2} + \beta U_{1} } \right) \right],
  & \alpha = 1.
\end{cases} \nonumber
 \end{equation}
where
\begin{equation}
B_{\alpha, \beta} = \frac{\tan^{-1} \left( \beta \tan \frac{\pi \alpha}{2}  \right)}{\alpha} \quad S_{\alpha, \beta} = \left( 1 + \beta^{2} \tan^{2} \frac{\pi \alpha}{2} \right)^{1 / 2 \alpha}. \nonumber
\end{equation}
Although it is hard to show for $\alpha$ and $\beta$, we can show that
\[
\mathbb{E}_{\theta} \left[ \left\{ \frac{\partial}{\partial \sigma} \tau_{\theta}(U) \right\}^{2} \right] = \mathbb{E}_{\theta} \left[ \left\{ \tau_{\alpha, \beta}(U) \right\}^{2} \right] =  \infty
\]
unless $\alpha = 2$. Therefore, it is not desirable to run the gradient ascent method for the process $\{ Y^{\epsilon}_{t} \}_{t \geq 1}$ with $Y^{\epsilon}_{t} = Y_{t} + \epsilon V_{t}$ since the variance of the gradient estimate will be infinite. Instead, we use the transformation $\psi = \tan^{-1}$, i.e.\ $ Y^{\epsilon}_{t} = \tan^{-1} (Y_{t}) + \epsilon V_{t}$ to make the gradient ascent method stable. One can indeed check that for the parameter $\sigma$ 
\[
\mathbb{E}_{\theta} \left[ \left\{ \frac{\partial}{\partial \sigma} \psi [ \tau_{\theta}(U) ] \right\}^{2} \right] = \mathbb{E}_{\theta} \left[ \left\{ \frac{\tau_{\alpha, \beta}(U)}{1 + \tau_{\theta}(U)^{2}} \right\}^{2} \right] < \infty
\]
We also verify numerically in Section \ref{sec: Numerical examples} that the gradients with respect to the other parameters $\alpha$, $\beta$ are stabilised with $\psi = \tan^{-1}$ (while we can show that $\mathbb{E}_{\theta} [ \left\{ \partial \tau_{\theta}(U) / \partial \mu \right\}^{2} ] = 1$).
\end{ex}

\section{Numerical examples} \label{sec: Numerical examples}

In this section we demonstrate the performance of the gradient ascent methods described in Section \ref{sec: Implementing ABC MLE with SMC} on the i.i.d.\ $\alpha$-stable and \textit{g-and-k} models as well as the stochastic volatility model with $\alpha$-stable returns.

\subsection{MLE for i.i.d.\ $\alpha$-stable random variables} \label{sec: MLE for alpha-stable distribution}

We first consider the problem of estimating the parameters of an $\alpha$-stable distribution $\mathcal{A}(\alpha, \beta, \mu, \sigma)$ (developed in Example \ref{ex: stability alpha stable}) from a sequence of i.i.d.\ samples. Several methods for estimating parameter values for stable distributions have been proposed, including a Bayesian approach based on ABC, see \citet{Peters_et_al_2011}. In this example we consider estimating these parameters using the online gradient ascent method to implement noisy ABC MLE. Since the only discontinuity in the transformation function $\tau_{\theta}$ for generating an $\alpha$-stable random variable is at $\alpha = 1$, we can safely use the gradient ascent method for estimating $\theta^{\ast}$ with $\alpha^{\ast}$ being not in the close vicinity of $1$. 

\begin{figure}[tbh]
\center{\includegraphics[scale = 0.62]{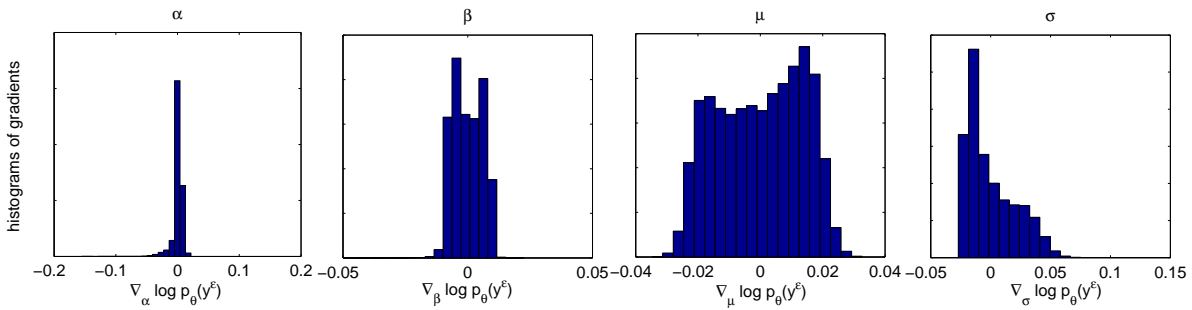}}
\caption{Histograms of estimates of $\nabla \log p_{\theta} (y^{\epsilon}_{i})$, $1 \leq i \leq 10^{5}$ computed at $\theta = (1.5, 0.5, 0, 0.5)$ where $y^{\epsilon}_{i}=\tan^{-1}(\hat{y}_{i})+0.1 v_{i}$, $\hat{y}_{i} \sim^{\text{i.i.d.}} \mathcal{A}(1.5, 0.5, 0, 0.5)$, $z_{i} \sim^{\text{i.i.d.}} \mathcal{N}(0,1)$.  
}
\label{fig: histogram of gradients with arctan}
\end{figure}

As recommended in Example \ref{ex: stability alpha stable}, we transform the observations using $\psi = \tan^{-1}$ for stability. In order to check, numerically, whether the transformation in \eqref{eq: stable noisy observation} with $\psi = \tan^{-1}$ stabilises the gradients, we can look at the empirical distribution of the Monte Carlo estimates of $\nabla \log p_{\theta}(Y^{\epsilon}_{i})$ after transforming the observations $Y_{i}$. For this purpose, we generate $10^{5}$ samples $\hat{y}_{i}$ from $\mathcal{A}(1.5, 0.5, 0, 0.5)$ and $v_{i}$ from $\kappa$ for $i = 1, \ldots, 10^{5}$, and for each sample we estimate $\nabla \log p_{\theta}(y^{\epsilon}_{i})$, where $y^{\epsilon}_{i} = \tan^{-1}(\hat{y}_{i}) + \epsilon v_{i}$, with $\epsilon = 0.1$, using self-normalised importance sampling with $N = 1000$ samples generated from $\nu_{\theta}$. Figure \ref{fig: histogram of gradients with arctan} shows the histograms of the Monte Carlo estimates of $\nabla \log p_{\theta}(y^{\epsilon}_{i})$ which confirms that the  transformation does stabilise the gradients.

\begin{figure}
\center{\includegraphics[scale = 0.58]{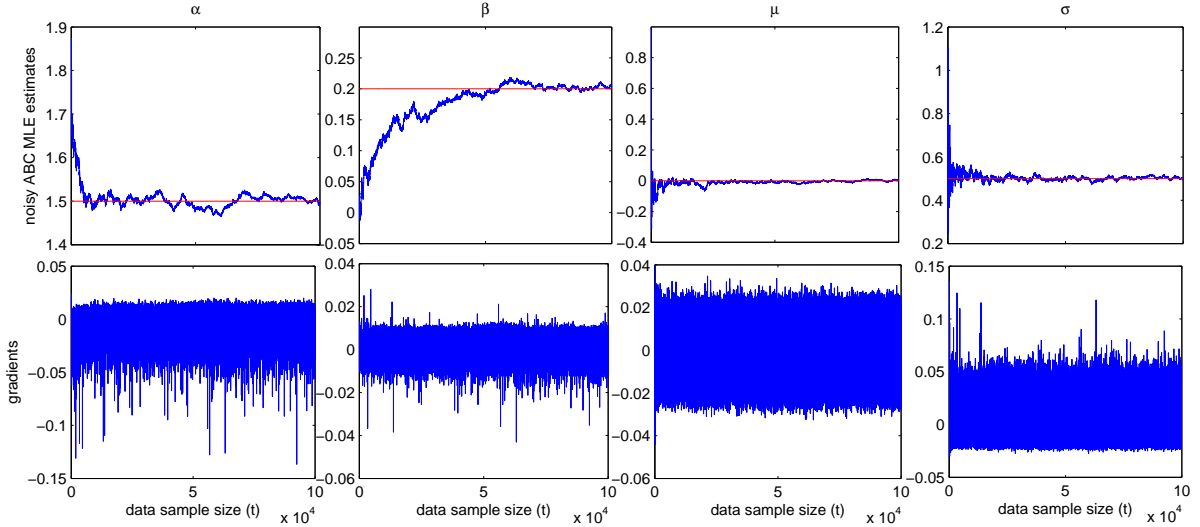}}
\caption{Online estimation of $\alpha$-stable parameters (top figure) from a sequence of i.i.d.\ random variables using online gradient ascent MLE and the corresponding online gradient estimates of the incremental likelihood (bottom figure). $\theta^{\ast} = (\alpha^{\ast}, \beta^{\ast}, \mu^{\ast}, \sigma^{\ast}) = (1.5, 0.2, 0, 0.5)$ is indicated with a horizontal line. At the bottom: Gradient of incremental likelihood for the $\alpha$-stable parameters}
\label{fig: estimation of alpha stable parameters}
\end{figure}

The outcome of online gradient ascent method to implement noisy ABC MLE for the same data set is shown in Figure \ref{fig: estimation of alpha stable parameters}. A trace plot of the sequence of gradient estimates (as $\theta$ is adjusted) is also shown as further confirmation of the stability of the estimated gradients.

\begin{figure}
\center{\includegraphics[scale = 0.60]{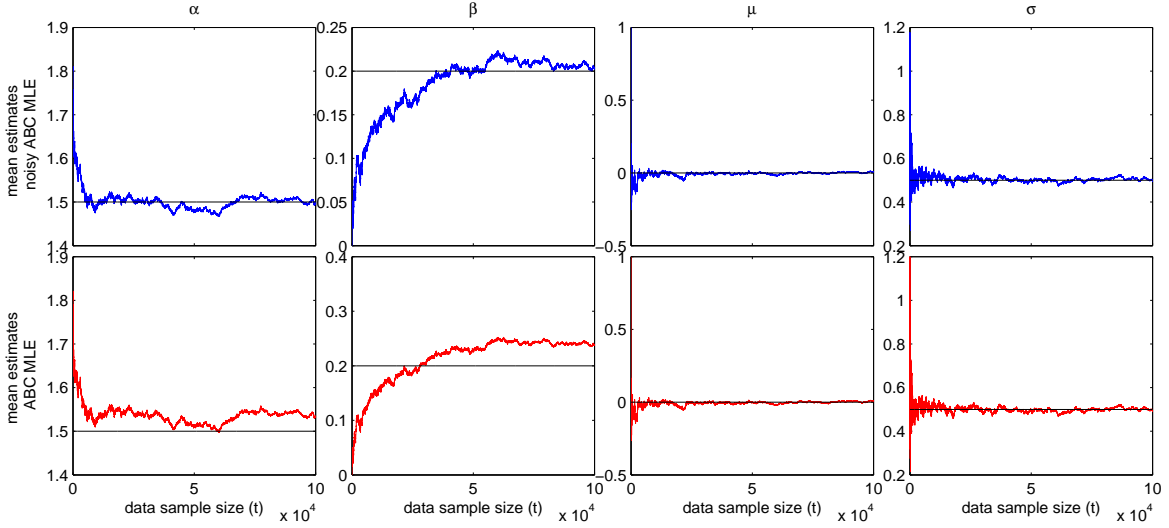}}
\caption{ABC MLE and noisy ABC MLE estimates of the parameters of the $\alpha$-stable distribution (averaged over 50 runs) using the online gradient ascent algorithm for the same data set. For noisy ABC MLE, a different noisy data sequence obtained from the original data set is used in each run. $\theta^{\ast} = (\alpha^{\ast}, \beta^{\ast}, \mu^{\ast}, \sigma^{\ast}) = (1.5, 0.2, 0, 0.5)$ is indicated with a horizontal line.}
\label{fig: comparison of noisy ABC MLE and noisy ABC MLE for alpha-stable}
\end{figure}

The next experiment contrasts the ABC MLE and noisy ABC MLE solutions for the same data set. The results in Figure \ref{fig: comparison of noisy ABC MLE and noisy ABC MLE for alpha-stable} compare the online $\theta^{\ast}$ estimates averaged over 50 independent runs for both algorithms. Each run used the same data set but a new realisation of particles. The outcome of this comparison is that ABC MLE yields biased estimates for the shape and skewness parameters $\alpha$ and $\beta$ whereas the bias is not present in noisy ABC MLE. 

\subsection{MLE for \textit{g-and-k} distribution} \label{sec: MLE for g-and-k distribution}
The \textit{g-and-k} distribution is defined by the following parameterised quantile (or inverse distribution) function $Q_{\theta}$
\begin{equation} \label{eq: quantile function for g-and-k}
Q_{\theta}(u) = F_{\theta}^{-1}(u) = A + B \left[ 1 + c \frac{1 - e^{-g \phi(u)}}{1 + e^{-g \phi(u)}} \right] \left( 1 + \phi(u)^{2} \right)^{k} \phi(u), \quad u \in (0, 1)
\end{equation}
where $\phi(u)$ is the $u$'th standard normal quantile. The parameters 
\[
\theta = (g, k, A, B) \in \Theta = \mathbb{R} \times (-0.5, \infty) \times \mathbb{R} \times [0, \infty)
\]
are the skewness, kurtosis, location and scale, and $c$ is usually fixed to $0.8$. Therefore one can generate from the \textit{g-and-k} distribution by first sampling $U \sim \text{Unif}(0, 1)$ and then returning $\tau_{\theta}(U) = Q_{\theta}(U)$ \citep{Rayner_and_MacGillivray_2002}.

Bayesian parameter estimation for the \textit{g-and-k} distribution using ABC was recently proposed in \cite{Fearnhead_and_Prangle_2012}. We consider online MLE for $\theta$ using the noisy ABC likelihood. Note that $Q_{\theta}$ in \eqref{eq: quantile function for g-and-k} is differentiable with respect to $\theta$ and so the gradient ascent method is applicable. To avoid gradients with very high variances resulting from the factor $\left( 1 + \phi(u)^{2} \right)^{k}$ in $Q_{\theta}$, similar to the case of $\alpha$-stable distribution, we transform the actual observations using $\psi = \tan^{-1}$ and add noise with $\epsilon = 0.1$. In our experiments it was noticed that our method performs better when the location parameter $A$ is closer to $0$, which must be a result of the non-linear behaviour of the transformation function $\tan^{-1}$. Therefore, whenever possible, it is suggested to estimate $A$ using some (possibly heuristic) method (such as using the mean of the first few samples) as a preprocessing step, subtract the heuristically estimated value $\hat{A}$ of $A$ from the samples, perform MLE on the (approximately) centred data, and then add back $\hat{A}$ to the estimated location obtained by the MLE algorithm. 

\begin{figure}[tbh]
\center{\includegraphics[scale = 0.64]{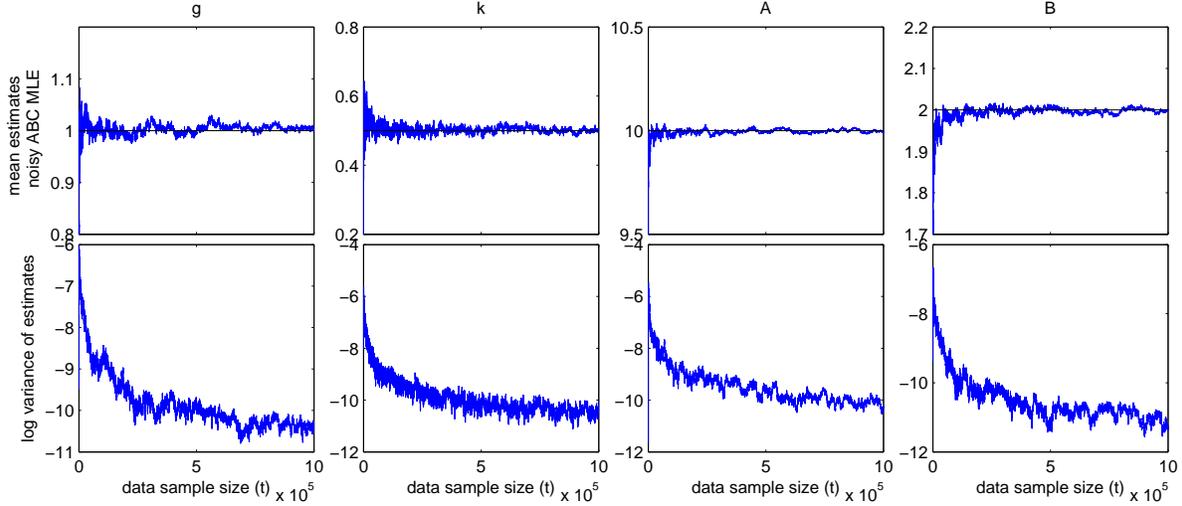}}
\caption{Mean and the variance (over 50 runs) of noisy ABC MLE estimates using the online gradient ascent algorithm. Same noisy data sequence used in each run. $\theta^{\ast} = (g^{\ast}, k^{\ast}, A^{\ast}, B^{\ast}) = (2, 0.5, 10, 2)$ indicated by horizontal lines.}
\label{fig: online estimation of g-and-k distribution parameters}
\end{figure}

Figure \ref{fig: online estimation of g-and-k distribution parameters} shows the results of online gradient ascent method \eqref{eq: recursive gradient MLE update} to implement noisy ABC MLE for estimating $\theta^{\ast} = (2, 0.5, 10, 2)$. In the figure we observe the mean and log-variance of 50 runs on the \textit{same} noisy transformed data sequence. (Therefore, the accuracy and the variance of the estimates correspond to the performance of the Monte Carlo approximation of the gradients $\nabla \log p_{\theta}(y^{\epsilon}_{i})$.) Self-normalised importance sampling is used with $N = 1000$ samples generated from $\nu_{\theta}$. From the results in Figure \ref{fig: online estimation of g-and-k distribution parameters}, we can see that the bias introduced by the finite number of particles is negligible for $N = 1000$ and that the variance of the algorithm reduces in time suggesting the convergence of the estimates in each run to essentially the true parameter values.
\begin{figure}[!tbh]
\center{\includegraphics[scale = 0.56]{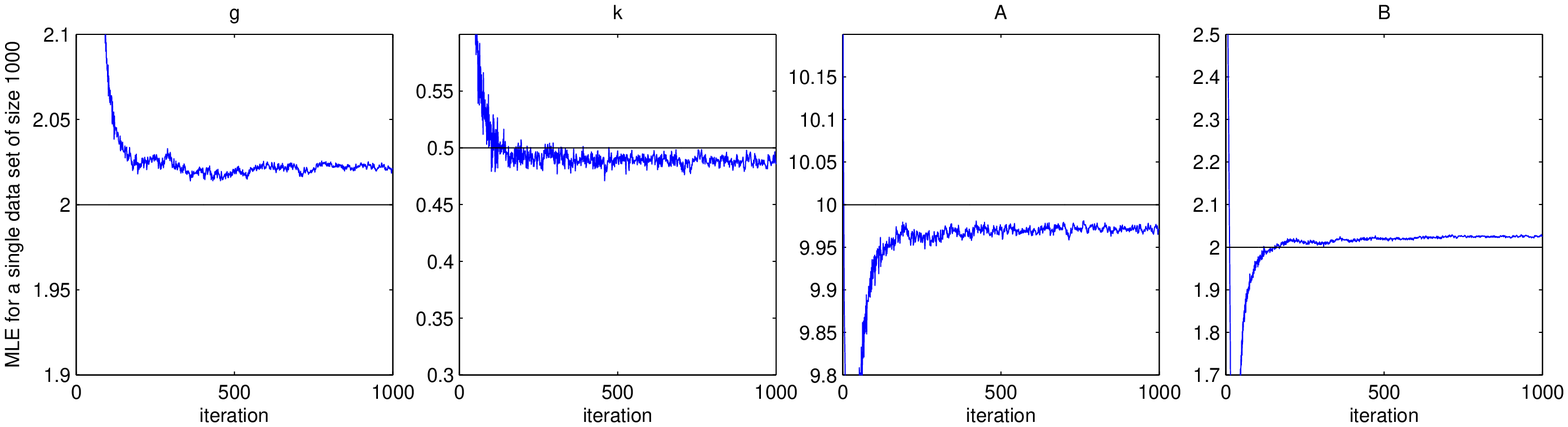}}
\vspace{-0.5in}
\center{\includegraphics[scale = 0.60]{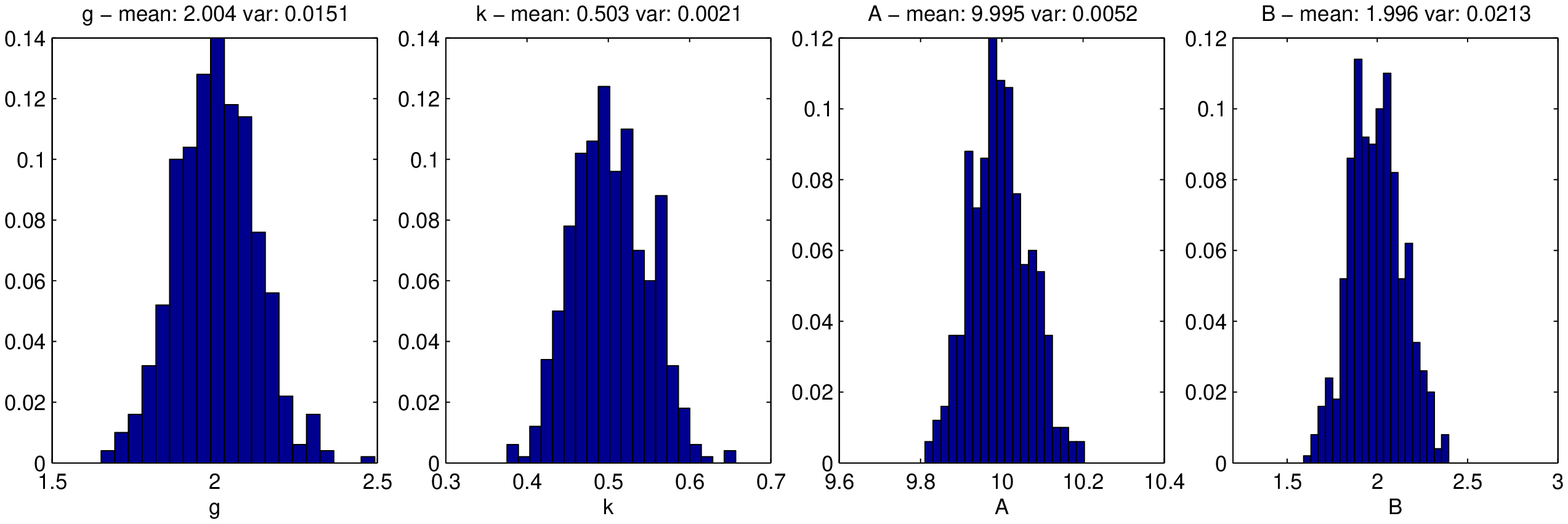}}
\caption{Top: noisy ABC MLE estimates of \textit{g-and-k} parameters from a sequence of 1000 i.i.d.\ random variables computed using the batch gradient ascent algorithm. $\theta^{\ast} = (g^{\ast}, k^{\ast}, A^{\ast}, B^{\ast}) = (2, 0.5, 10, 2)$ indicated by horizontal lines. Bottom: Empirical distributions of the estimates over 500 data sets.}
\label{fig: batch estimation of g-and-k distribution parameters}
\end{figure}

The next experiment shows how the noisy ABC MLE can be implemented with the batch gradient ascent method \eqref{eq: gradient MLE update} when the data set is too small for the online method to converge. A detailed study of MLE for \textit{g-and-k} distribution can be found in \citet{Rayner_and_MacGillivray_2002} where MLE methods based on numerical approximation of the likelihood itself are investigated. We generated 500 data sets of size $n = 1000$ from the same \textit{g-and-k} distribution with $\theta^{\ast} = (2, 0.5, 10, 2)$ and executed the batch gradient ascent method with $\epsilon = 0.1$ on each data set. Again, self-normalised importance sampling is used with $N = 1000$ samples. The upper half of Figure \ref{fig: batch estimation of g-and-k distribution parameters} shows the estimation results with noisy ABC MLE versus number of iterations for a single data set. Note that for short data sets, $\theta^{\ast}$ is usually not the true maximum likelihood solution. The lower half of Figure \ref{fig: batch estimation of g-and-k distribution parameters} shows the distributions (histograms over 20 bins) of the converged maximum likelihood solution for $\theta^{\ast}$. The mean and variance of the estimates for $(g, k, A, B)$ are $(2.004, 0.503, 9.995, 1.996)$ and $(0.0151, 0.0021, 0.0052, 0.0213)$ respectively. Comparable values for these moments at this particular $\theta^{\ast}$ and data size $n$ were also obtained in \citet[Table 3]{Rayner_and_MacGillivray_2002}.
  
\subsection{The stochastic volatility model with symmetric $\alpha$-stable returns} \label{The stochastic volatility model with symmetric alpha-stable returns}
\begin{figure} 
\center{\includegraphics[scale = 0.65]{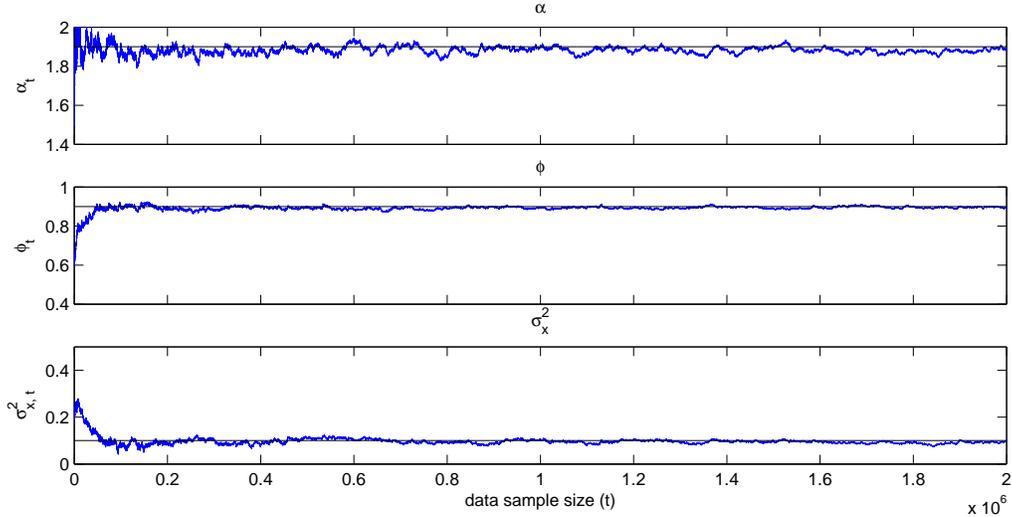}}
\caption{Online estimation of SV$\alpha$R parameters using online gradient ascent algorithm to implement noisy ABC MLE. $\theta^{\ast} = (\alpha^{\ast}, \phi^{\ast}, \sigma_{x}^{2, \ast}) = (1.9, 0.9, 0.1)$ indicated with horizontal lines.}
\label{fig: estimation of stochastic volatility model parameters with gradient ascent}
\end{figure}

The stochastic volatility model with $\alpha$-stable returns (SV$\alpha$R) is a financial data model \citep{Lombardi_and_Calzolari_2009}. The hidden process $\{ X_{t} \}_{t \geq 1}$ represents the log-volatility in time whereas the observation process $\{ Y_{t} \}_{t \geq 1}$ is the log return values. The model for $\{ X_{t}, Y_{t} \}_{t \geq 1}$ with parameters $\theta = (\alpha, \phi, \sigma_{x}^{2})$ is:
\begin{align*}
X_{t} = \phi X_{t-1} + S_{t}, \quad S_{t} \sim \mathcal{N}(0, \sigma_{x}^{2}), \quad Y_{t} = \exp(X_{t}/2) W_{t}, \quad W_{t} \sim \mathcal{A} (\alpha, 0, 0, 1).
\end{align*}
This model is an alternative to the stochastic volatility model with Gaussian returns to account for an observed series which is heavy-tailed and displays outliers. For more discussion on the model as well as a review of methods for estimating the static parameters of such models, see \citet{Lombardi_and_Calzolari_2009} and the references therein. These existing methods for parameter estimation in SV$\alpha$R are batch and suitable for only short data sequences. We simulated a scenario where a very long data sequence generated from this model with $\theta^{\ast} = (1.9, 0.9, 0.1)$ is being received sequentially. We used online gradient ascent method \eqref{eq: recursive gradient MLE update} to find the noisy ABC MLE solution for this data sequence, where the $\mathcal{O}(N^{2})$ method \citep[Algorithm 2]{Poyiadjis_et_al_2011} with $N = 500$ particles was used to estimate \eqref{eq: gradient of the incremental log-likelihood}. Again, we transform the actual observations with the function $\psi = \tan^{-1}$ and then add noise. Figure \ref{fig: estimation of stochastic volatility model parameters with gradient ascent} shows the online estimates of $\theta^{\ast}$ for $2 \times 10^{6}$ data samples. The estimates seem to converge after around $5 \times 10^{5}$ samples and are accurate.

\subsection{Offline noisy ABC MLE for real data} \label{sec: Batch gradient MLE for real data}

We now consider a real data experiment, where the data are the daily GBP-DEM exchange rates between $01.01.1987$ to $31.12.1995$ containing $3287$ samples $o_{1}, \ldots, o_{3287}$; these data are considered in \citet{Lombardi_and_Calzolari_2009}. Log-returns $r_{1:3286}$ are obtained by $r_{t} = 100 \log (o_{t + 1}/o_{t})$, $1 \leq t \leq 3286$. The observations, $\hat{y}_{1:3285}$, are the residuals of the AR(1) process that is fitted to $r_{1:3286}$. (We used the same model and data set as \citet{Lombardi_and_Calzolari_2009} in order to compare our results with theirs). The SV$\alpha$R model  above is assumed for $\hat{y}_{1:n}$, where the hidden process has an extra parameter $\delta$:
\[
X_{t} = \phi X_{t-1} + \delta + S_{t}, \quad S_{t} \sim \mathcal{N}(0, \sigma_{x}^{2}),
\]
hence $\theta = (\alpha, \phi, \sigma_{x}^{2}, \delta)$.

\begin{figure} 
\center{\includegraphics[scale = 0.72]{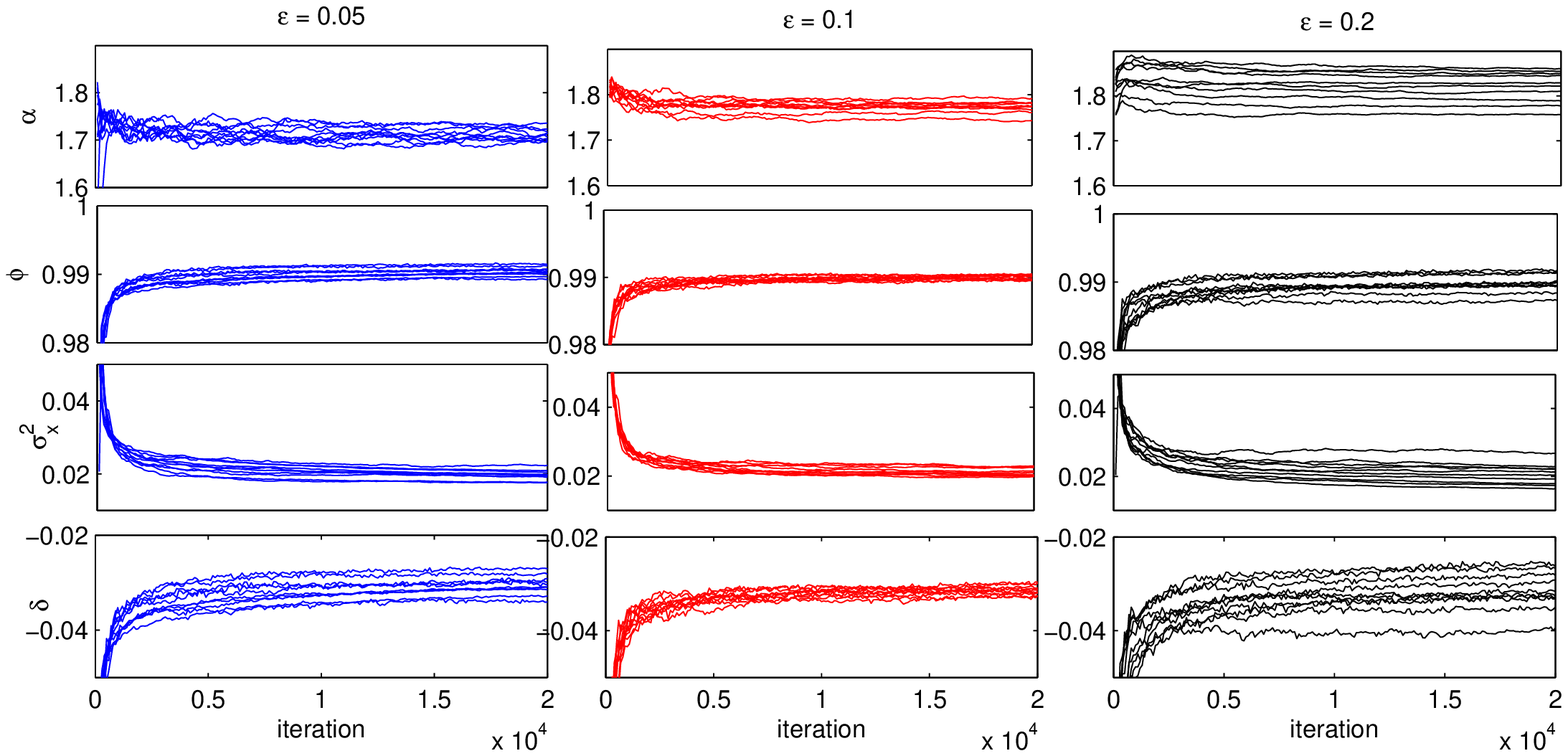}}
\center{\includegraphics[scale = 0.60]{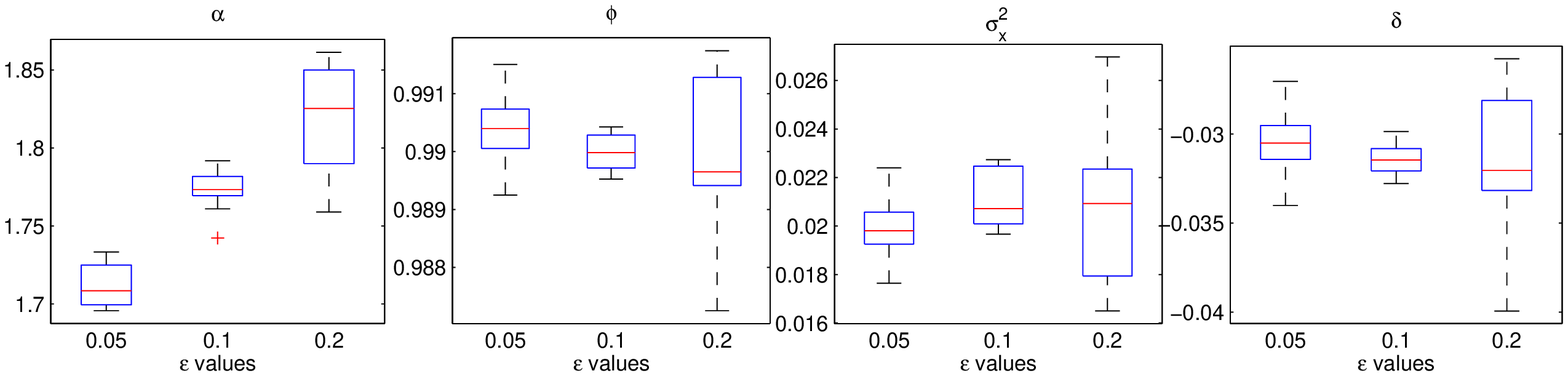}}
\caption{Top: results for noisy ABC MLE implemented by the $\mathcal{O}(N)$ batch gradient MLE algorithm for three different values of $\epsilon$. Estimates vs number of iterations for different noisy data sets are superimposed for the same value of $\epsilon$. Bottom: box-plots of the batch ABC MLE estimates vs $\epsilon$. The box-plot for each $\epsilon$ was created from $\theta_{\epsilon}^{(1)}, \ldots, \theta_{\epsilon}^{(10)}$, the converged estimates obtained from the trace plots at the top.}
\label{fig: batch MLE trajectories and box-plots for real data}
\end{figure}

We implemented noisy ABC MLE using batch gradient ascent \eqref{eq: gradient MLE update} with the $\mathcal{O}(N)$ method \citep[Algorithm 1]{Poyiadjis_et_al_2011} with $N = 2000$ particles to approximate \eqref{eq:score}. To measure the variability of the estimates as a function of the realisation of added noise and the $\epsilon$ value, we repeated the estimation with $\epsilon = 0.05$, $\epsilon = 0.1$, and $\epsilon = 0.2$, separately, where for each $\epsilon$ we ran the method with $10$ different added noise realisations. For all runs, we terminated the batch gradient ascent algorithm after $20000$ iterations. $N = 2000$ particles were used to evaluate the gradients at each iteration. Figure \ref{fig: batch MLE trajectories and box-plots for real data} (top) shows the estimates versus number of iterations, where the trajectories for different noisy data sets for the same value of $\epsilon$ are superimposed. Also, the bottom part of Figure \ref{fig: batch MLE trajectories and box-plots for real data} shows the box-plots of the estimates of $\theta^{\ast}$ for different $\epsilon$ values, where the box-plots for each $\epsilon$ were created from the converged estimates of $\theta^{\ast}$ (the average of the estimates at the last 1000 iterations) obtained from 10 different noisy data sets generated using that value of $\epsilon$. For the ease of explanation, we will denote them as 
\begin{equation} \label{eq: estimates for different epsilon values}
\theta_{0.05}^{(1)}, \ldots, \theta_{0.05}^{(10)}; \theta_{0.1}^{(1)}, \ldots, \theta_{0.1}^{(10)}; \theta_{0.2}^{(1)}, \ldots, \theta_{0.2}^{(10)}, 
\end{equation}
where $\theta_{\epsilon}^{(i)}$ is the converged estimate obtained from the $i$'th noisy data set that was generated using $\epsilon$. 

Figures \ref{fig: batch MLE trajectories and box-plots for real data} suggests a trade off between accuracy in the estimates and computational efficiency in the following sense. A smaller value of $\epsilon$ is expected yield less biased estimates (with respect to the maximiser of the true likelihood of the real data) with less variance (with respect to the added noise) provided that the maximisation $\arg \max_{\theta \in \Theta} p_{\theta}(Y^{\epsilon}_{1:n} = y^{\epsilon}_{1:n})$ is performed exactly, that is with infinitely many $N$ and infinitely many number of parameter updates. On the other hand, smaller $\epsilon$ results in the decrease of the effective sample size in the SMC algorithm and hence increases the variance of the SMC estimate of the gradient of the log likelihood. The effect of this on our results is the larger variance in the estimates obtained with $\epsilon = 0.05$ compared to those obtained with $\epsilon = 0.1$ (which would eventually be smaller if the maximisation were performed exactly). In conclusion, for a fixed batch data size and a given amount of computational resource, one must optimise the trade off between the (average) accuracy and the variability in the estimates, for which the effective sample size of the particles could be used as a rule of thumb.

\begin{figure}
\center{\includegraphics[scale = 0.65]{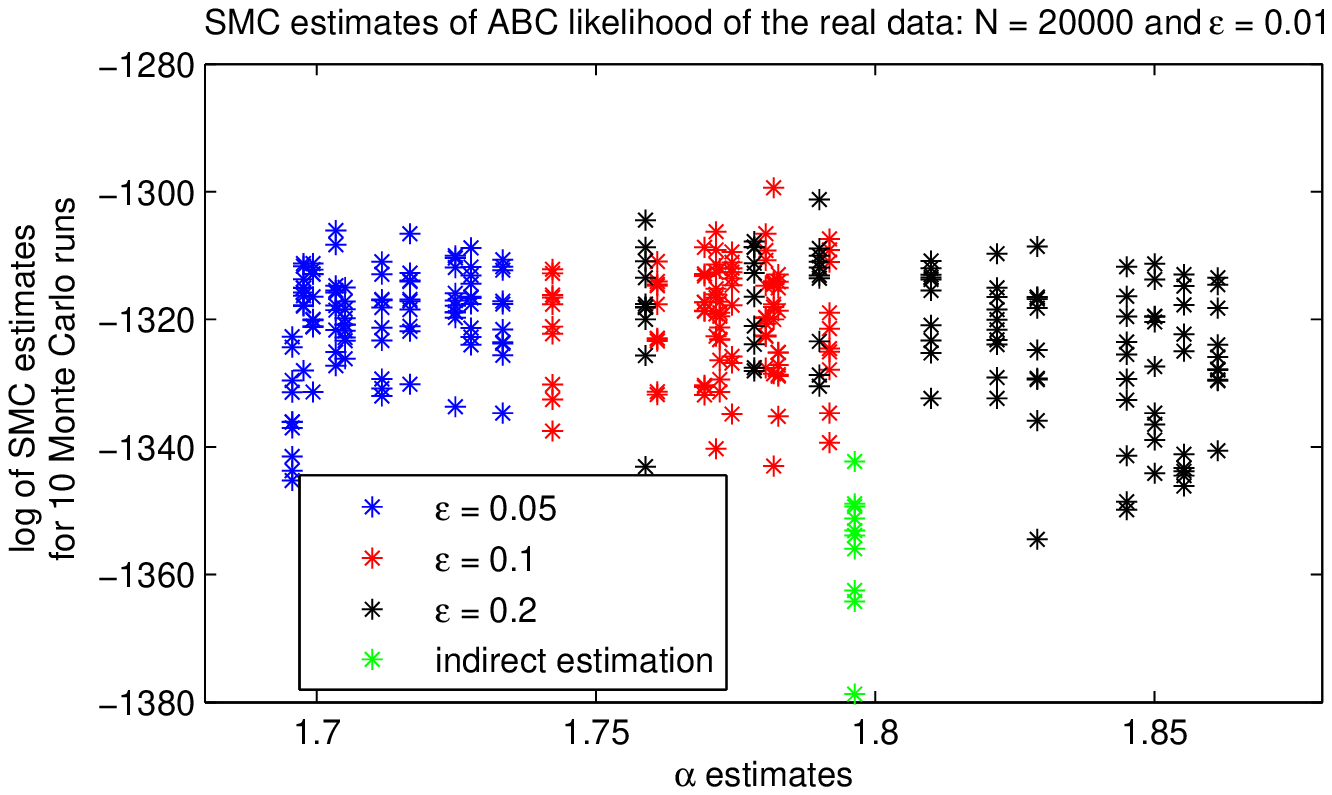}\includegraphics[scale = 0.65]{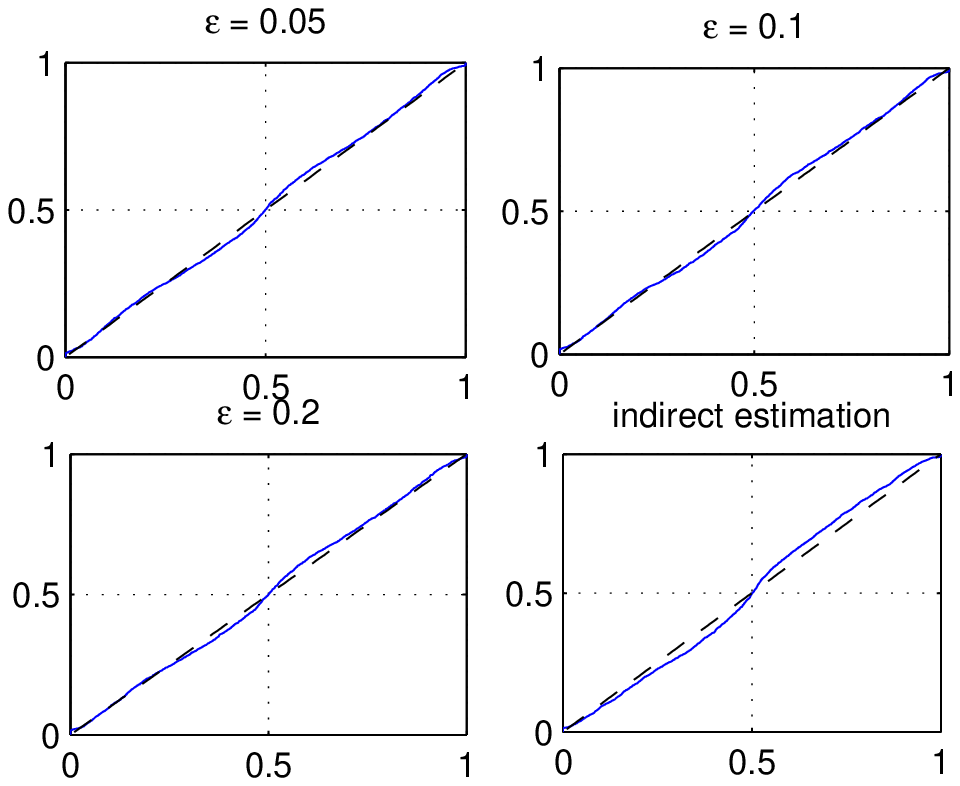}}
\caption{Left: Logarithm of the 10 different SMC estimates (with $N = 20000$) of $p_{\theta}(Y_{1:n}^{\epsilon} = \hat{y}_{1:n})$, $\epsilon = 0.01$. Each colour represents a different $\epsilon$ value which was used to obtain the noisy data sets and the ABC MLE estimates from them. For the blue, red, and black points, their horizontal axis locations correspond to the $\alpha$ components of $\theta_{\epsilon}^{(1)}, \ldots, \theta_{\epsilon}^{(10)}$ for $\epsilon = 0.05$, $\epsilon = 0.1$, and $\epsilon = 0.2$, respectively. Similarly, the horizontal axis location of the black points is the $\alpha$ component of the estimate of $\theta^{\ast}$ obtained using the indirect estimation method. Right: Empirical cumulative distribution plots for model checking: for each $\epsilon$ value, $\theta$ is taken to be the mean of $\theta_{\epsilon}^{(1)}, \ldots, \theta_{\epsilon}^{(10)}$.}
\label{fig: comparison of SMC likelihood estimates for real data}
\end{figure}

\citet{Lombardi_and_Calzolari_2009} fitted the same model to the same data set using the indirect estimation method and their estimates of $\theta^{\ast}$ was $\theta_{\text{ind}} := (1.7963, 0.9938, 0.0940^2, -0.0076)$, which is slightly different to our results. Both ours and their method aim for the maximum likelihood solution, which suggests that it would be sensible to compare the likelihood of the true data sequence for the estimates of $\theta^{\ast}$ obtained from both methods. However, this is not possible since neither $p_{\theta}(Y_{1:n} = \hat{y}_{1:n})$ nor an unbiased Monte Carlo estimator of it is available. Instead, we compared the unbiased SMC estimates of the ABC likelihoods $p_{\theta}(Y_{1:n}^{\epsilon} = \hat{y}_{1:n})$ using an $\epsilon$ small enough to make the effect of model mismatch negligible (see the discussion of model mismatch error in Section 2) for comparison and $N$ large enough to ensure that the variability of the SMC estimate of the likelihood across the particle realisations is not too much; for these reasons we chose $\epsilon = 0.01$ and $N = 20000$. (See Appendix for the details of the implementation.) The left hand side of Figure \ref{fig: comparison of SMC likelihood estimates for real data} shows the logarithms of the 10 independent SMC estimates of $p_{\theta}(Y_{1:n}^{\epsilon} = \hat{y}_{1:n})$ calculated at the value of each estimate in \eqref{eq: estimates for different epsilon values}. For comparison, the results are shown with 10 independent SMC estimates of $p_{\theta}(Y_{1:n}^{\epsilon} = \hat{y}_{1:n})$ at $\theta = \theta_{\text{ind}}$. The figure shows that noisy ABC MLE has improved the results of \citet{Lombardi_and_Calzolari_2009} for all values of $\epsilon$ that we used, in the sense that almost all the estimates resulting from the ABC MLE method yields a higher likelihood of the data set to which the model is fitted.

Finally, we perform a simple model check for by considering the conditional cumulative distribution functions
\[
F_{\theta, t}(\hat{y}_{t}; \hat{y}_{1:t-1}) := P_{\theta}(Y_{t} \leq \hat{y}_{t} | Y_{1:t-1} = \hat{y}_{1:t-1}), \quad t = 1, \ldots, n.
\]
at the values of $\theta^{\ast}$ estimated using noisy ABC MLE and the indirect estimation method in \citet{Lombardi_and_Calzolari_2009}. Since $\{ F_{\theta, t}(Y_{t}; Y_{1:t-1}) \}_{1 \leq t \leq n}$ are i.i.d.\ uniform random variables on $[0, 1]$ \citep{Diebold_et_al_1998}, we expect the probability plot (for the uniform distribution) of the population $\{ F_{\theta, t}(\hat{y}_{t}; \hat{y}_{1:t-1}) \}_{1 \leq t \leq n}$ to approximate the $y = x$ line under the hypothesis that $\hat{y}_{1:n}$ is generated from the SV$\alpha$R model $\{ X_{t}, Y_{t} \}_{t \geq 1}$. However, we are unable to perform these calculations for the original HMM due to the intractability of $g_{\theta}(y |x)$. Instead, we use the modified HMM $\{ (X_{t}, Y_{t}), Y^{\epsilon}_{t} \}_{t \geq 1}$ but with $\epsilon$ small enough for one to neglect the difference between the two models (as in the previous experiment). The probability plots at the right hand side of Figure \ref{fig: comparison of SMC likelihood estimates for real data} were generated from the SMC estimates of
\[
F_{\epsilon, \theta, t}(\hat{y}_{t}; \hat{y}_{1:t-1}) := P_{\theta}(Y^{\epsilon}_{t} \leq \hat{y}_{t} | Y^{\epsilon}_{1:t-1} = \hat{y}_{1:t-1}), \quad t = 1, \ldots, n,
\]
(see Appendix for the details), with $\epsilon = 0.01$ and $N = 20000$, for four different values of $\theta$: the first three are the means of $\theta_{\epsilon}^{(1)}, \ldots, \theta_{\epsilon}^{(10)}$ for $\epsilon = 0.05$, $\epsilon = 0.1$ and $\epsilon = 0.2$, respectively, and the fourth one is $\theta_{\text{ind}}$. The probability plots are all close to the $y = x$ line which justifies the SV$\alpha$R model; they also indicate that there is more agreement between the SV$\alpha$R model and the data when $\theta$ is the noisy ABC MLE solution than when it is the maximum likelihood solution of the indirect estimation method.

\section{Discussion} \label{sec: Discussion}

In this paper, we have presented SMC implementations of MLE for HMMs with an intractable observation density. We showed how SMC versions of both batch and online gradient ascent algorithms can be used to implement noisy ABC MLE and how a further transformation of the data can stabilise the variance of the SMC gradient estimate. We have shown that SMC implementations of the methodology in \citet{Dean_et_al_2011} is practical and yields convergent and accurate estimates of $\theta^{\ast}$ even when the exact procedures in \citet{Dean_et_al_2011} are replaced by their SMC counterparts.

\subsection{Other MLE methods for HMMs with an intractable density} \label{sec: Other MLE methods for HMMs with an intractable density}
Although not as general as the gradient ascent MLE approach, the expectation-maximisation (EM) algorithm may be available for some models, at least for a part of the parameters in $\theta$, if the joint density $p_{\theta}(z_{1:n}, y^{\epsilon}_{1:n}) $ belongs to an exponential family. Both $\mathcal{O}(N)$ and $\mathcal{O}(N^{2})$ batch and online EM algorithms can be devised using SMC; details of such algorithms can be found in \citet{Cappe_2009a} and \citet{Del_Moral_et_al_2009}.

There are other gradient MLE methods in the literature that are available for implementing noisy ABC MLE and we have discussed the technique of \citet{Ehrlich_et_al_2013} in the introduction. One advantage of their finite difference method is that it is essentially a gradient free technique as it bypasses having to calculate the derivatives with respect to $\theta$ of the state transition and observation densities of the HMM and thus can cope, without modification, with an intractable state transition density. Another gradient based method that uses SMC to approximate the gradient of the log-likelihood without the need to calculate the derivatives of the HMMs densities is the iterated filtering algorithm of \citet{Ionides_et_al_2011}. In particular, one can use iterated filtering for $\{ (X_{t}, Y_{t}), Y^{\epsilon}_{t} \}_{t \geq 1}$ or $\{ (X_{t}, U_{t}), Y^{\epsilon}_{t} \}_{t \geq 1}$ in order to estimate $\nabla \log p_{\theta}(y^{\epsilon}_{1:n})$. However, the method does not have an extension to online estimation. Another downside is that the algorithm requires an increasing number of particles versus iteration for convergence.

\citet{Coquelin_et_al_2009} study a HMM with a tractable observation density $g_{\theta}(y | x)$ but an intractable state transition density $f_{\theta}(x' | x)$. Assume one can generate from $f_{\theta}(\cdot | x)$ by sampling $U$ from $\mu_{\theta}(\cdot | x)$ and using a differentiable function $F_{\theta}: \mathcal{X} \times \mathcal{U} \rightarrow \mathcal{X}$  such that $F_{\theta}(U, x) \sim f_{\theta}(\cdot | x)$. The gradient of the log likelihood in such HMMs can be estimated using the infinitesimal perturbation analysis (IPA) approach proposed in \citet{Coquelin_et_al_2009}, provided that $\mu_{\theta}(\cdot | x)$, $F_{\theta}(u, x)$, and $g_{\theta}(y | x)$ are differentiable with respect to $\theta$ as well as the state variable $x$. We can straightforwardly adopt the IPA approach with our noisy ABC MLE to deal with a fully intractable model, where both the state transition and the observation densities are intractable. However, IPA is a path space method and suffers from particle degeneracy. This will lead to the variance of the estimate of the score in \eqref{eq:score} increasing quadratically in time like the $\mathcal{O}(N)$ method in \citet{Poyiadjis_et_al_2011}. As the authors mention, fixed-lag smoothing could be use to control this variance growth but at the cost of a small bias.

Static parameter estimation for HMMs with intractable state and observation densities have been addressed in a Bayesian context by \citet{Campillo_and_Rossi_2009}. \citet{Campillo_and_Rossi_2009} utilise the so called convolution particle filter, which uses ideas from kernel density estimation to replace the intractable densities needed for the weight evaluation in the particle filter with their kernel estimates, to sequentially estimate the posterior distribution of $\theta^{\ast}$. While an SMC based Bayesian approach can potentially produce good estimates of $\theta^{\ast}$ for short data lengths, at least for tractable models where standard particle methods apply, particle degeneracy does bias the estimation results for long data sets \citep{Andrieu_et_al_2005, Kantas_et_al_2009}. In contrast our methods do give rise to practically consistent estimators as our numerical results indicate.

Finally, we remark that MLE using ABC is studied in the recent work \citet{Rubio_and_Johansen_2013}, but in a non-HMM setting where the likelihood of data $\hat{y}$ given $\theta$ is intractable. The authors form a kernel density estimate of the likelihood from $\theta$ samples drawn from the ABC posterior distribution. They propose maximising the kernel density estimate as an approximation to MLE. Unlike \citet{Rubio_and_Johansen_2013}, we consider the HMM setting and our methods do not need samples of $\theta$.

\section*{Acknowledgement}
S.S.\ Singh and T.\ Dean's research was funded by the Engineering and Physical Sciences Research Council (EP/G037590/1) whose support is gratefully acknowledged. A.\ Jasra was supported by an MOE Singapore grant and is also affiliated with the Risk Management Institute at the National University of Singapore.

\footnotesize{
\linespread{1.0}
\bibliographystyle{apalike} 
\bibliography{myrefs_thesis}
}

\appendix

\normalsize
\section*{Appendix} 

\begin{alg} \textbf{SMC for estimating $p_{\theta}(Y^{\epsilon}_{1:n} = \hat{y}_{1:n})$ and $\{ F_{\epsilon, \theta, t}(\hat{y}_{t} | \hat{y}_{1:t-1}) \}_{1 \leq t \leq n}$}\\

Begin with $p_{\theta}(\hat{y}_{0}) = 1$. For $t = 1, \ldots, n$,
\begin{itemize}
\item \emph{Prediction:}  for $i = 1, \ldots, N$, sample $z_{t}^{(i)} = (x_{t}^{(i)}, u_{t}^{(i)})$ as follows:
\begin{itemize}
\item If $t = 1$, sample $x_{1}^{(i)} \sim \eta_{\theta}(\cdot)$, $u_{1}^{(i)} \sim \nu_{\theta}(\cdot | x_{1}^{(i)})$ 
\item If $t > 1$, sample $x_{t}^{(i)} \sim f_{\theta}(\cdot | \bar{x}_{t-1}^{(i)})$, $u_{t}^{(i)} \sim \nu_{\theta}(\cdot | x_{t}^{(i)})$. 
\end{itemize}

\item \emph{Weighting:} for $i = 1, \ldots, N$, calculate the unnormalised weights $w_{t}^{(i)} = h^{\epsilon}(\hat{y}_{t} | z_{t}^{(i)})$
\item \emph{Likelihood estimate:} Update the likelihood estimate by $p_{\theta}^{N}(\hat{y}_{1:t}) = p_{\theta}^{N}(\hat{y}_{1:t-1}) \frac{1}{N} \sum_{i =1}^{N} w_{t}^{(i)}$.
\item \emph{conditional cumulative distribution function:} Calculate
\[
F_{\epsilon, \theta, t}^{N}(\hat{y}_{t}; \hat{y}_{1:t-1}) = \frac{\sum_{i = 1}^{N} w_{t}^{(i)} \int_{\infty}^{\hat{y}_{t}}h^{\epsilon}(y | z_{t}^{(i)}) dy}{\sum_{j = 1}^{N} w_{t}^{(j)}}.
\]
\item \emph{Resampling:} Sample $\{ \bar{x}_{t}^{(i)} \}_{1 \leq i \leq N}$ from $\{ x_{t}^{(i)} \}_{1 \leq i \leq N}$ using the weights $\{ w_{t}^{(i)} \}_{i = 1, \ldots, N}$.
\end{itemize}
\end{alg}

\end{document}